\documentclass[aps,superscriptaddress,amsmath,amssymb,twocolumn,amsfonts,floatfix,nofootinbib,longbibliography]{revtex4-1}
\usepackage{graphicx}
\usepackage{color}
\graphicspath{{Pictures/}}
\usepackage{hyperref}
\usepackage{soul}
\usepackage{natbib}
\hypersetup{pdfnewwindow=true, colorlinks=true, linkcolor=magenta, anchorcolor=blue, citecolor=red, filecolor=blue, menucolor=blue, urlcolor=magenta}
\usepackage{cleveref}
\usepackage{color}
\usepackage{float}
\usepackage{soul}
\usepackage{braket}
\usepackage{enumitem}
\usepackage{marginnote}
\usepackage[normalem]{ulem}

\newcommand{\SW}[1] {{\color{black}#1}}

\newcommand{\va}{\ensuremath{\mathbf{a}}}

\newcommand{\vk}{\ensuremath{\mathbf{k}}}
\begin{document}
  
\title{Higher-order topological corner and bond-localized modes in magnonic insulators}
    
\author{Sayak Bhowmik}
\email{sayak.bhowmik@iopb.res.in}
\affiliation{Institute of Physics, Sachivalaya Marg, Bhubaneswar, Orissa 751005, India}
\affiliation{Homi Bhabha National Institute, Training School Complex, Anushakti Nagar, Mumbai 400094, India}
\author{Saikat Banerjee}
\email{saikat.banerjee@rutgers.edu}
\affiliation{Theoretical Division, T-4, Los Alamos National Laboratory, Los Alamos, New Mexico 87545, USA}
\affiliation{Center for Materials Theory, Rutgers University, Piscataway, New Jersey, 08854, USA}
\author{Arijit Saha}
\email{arijit@iopb.res.in}
\affiliation{Institute of Physics, Sachivalaya Marg, Bhubaneswar, Orissa 751005, India}
\affiliation{Homi Bhabha National Institute, Training School Complex, Anushakti Nagar, Mumbai 400094, India}
%\date{\today}
    
%------------------
\begin{abstract}
We theoretically investigate a two-dimensional decorated honeycomb lattice framework to realize a second-order topological magnon insulator (SOTMI) phase featuring distinct corner-localized modes. Our study emphasizes the pivotal role of spin-magnon mapping in characterizing bosonic topological properties, which exhibit differences from their fermionic counterparts. We employ a symmetry indicator topological invariant to identify and characterize this SOTMI phase, particularly for systems respecting time-reversal and ${\sf{C}}_6$ rotational symmetry. Using a spin model defined on a honeycomb lattice geometry, we demonstrate that introducing ``\textit{kekul\'e}'' type distortions yields a topological phase. In contrast, ``\textit{anti-kekul\'e}'' distortions result in a non-topological magnonic phase. The presence of kekul\'e distortions manifests in two distinct topologically protected bosonic corner modes - an \textit{intrinsic} and a \textit{pseudo}, based on the specific edge terminations. On the other hand, anti-kekul\'e distortions give rise to \SW{\textit{Tamm/Shockley}} type bond-localized boundary modes, which are non-topological and reliant on particular edge termination. We further investigate the effects of random out-of-plane exchange anisotropy disorder on the robustness of these bosonic corner modes. The distinction between SOTMIs and their fermionic counterparts arises due to the system-specific magnonic onsite energies, a crucial feature often overlooked in prior literature. Our study unveils exciting prospects for engineering higher-order topological phases in magnon systems and enhances our understanding of their unique behavior within decorated honeycomb lattices.
\end{abstract}
%------------------
    
\maketitle
   
{\textcolor{blue}{\textit{Introduction}}} \--- Since the discovery of graphene~\cite{RevModPhys.81.109,RevModPhys.84.1067} and the advent of topological insulators~\cite{RevModPhys.82.3045}, non-Bravais lattices have become a central platform for exploring exotic quantum phenomena in contemporary condensed matter physics. On one front, extensive research has been dedicated to the topological classification of non-interacting systems, encompassing both symmorphic~\cite{Parameswaran2013,PhysRevX.6.031003} and non-symmorphic crystalline structures~\cite{PhysRevB.90.085304,PhysRevB.91.161105,PhysRevB.103.235125}, over the past decade. On the other hand, the focus has shifted towards extending these concepts into the realm of correlated electron systems, notably in topological Mott~\cite{Maciejko2015,Hasan2011,Rachel_2018} and Kondo insulators~\cite{PhysRevLett.104.106408,Dzero_2016}. 
   
Among the various non-Bravais lattices, honeycomb lattice is one of the \SW{simplest examples because of its abundance in various quantum materials,} \SW{that accomodates} intricate multiband physics without strong interactions. While it is renowned for its fermionic Dirac excitations and topological quasiparticles, recent research has extended its significance to encompass bosonic systems featuring novel topological band structures. These advances have spurred the exploration of diverse bosonic topological materials, spanning photonic crystals~\cite{PhysRevLett.100.013904,PhysRevB.97.035442}, plasmonic systems~\cite{PhysRevLett.110.106801,Wu2017}, arrays of superconducting grains~\cite{PhysRevB.93.134502}, and magnetic structures~\cite{PhysRevB.94.075401,PhysRevX.8.011010,Mcclarty2022} etc. and opened up a new and compelling avenue for scientific inquiry. Recently, higher-order topological insulators (HOTI) have gained significant attention as a novel and intriguing class of systems\--- extended for both fermionic~\cite{Benalcazar2017, Frank2018, benalcazarprb2017,Langbehn2017, Franca2018,wang2018higher,Geier2018,Roy2019, Ezawakagome,RoyGHOTI2019,Trifunovic2019,Khalaf2018, BiyeXie2021,trifunovic2021higher, SchindlerDirac2020} and bosonic platform~\cite{PhysRevLett.125.207204, PhysRevB.104.024406, Li2019,PhysRevB.98.235102,PhysRevB.99.235132,PhysRevB.102.041126}. In a $d$-dimensional space, an $n^{\rm{th}}$ order HOTI is characterized by the presence of $(d-n)$ dimensional boundary modes. More specifically, two-dimensional (2D) second-order topological magnon insulators (SOTMI) exhibit a distinct presence of a finite number of corner modes. Previous theoretical investigations into various systems ranging from the breathing kagome and square lattices~\cite{Sil_2020,PhysRevB.101.184404} to skyrmion crystals~\cite{PhysRevLett.125.207204} and twisted bi-layer honeycomb networks~\cite{PhysRevB.107.L020404} have demonstrated the existence of such topologically protected magnonic corner modes. An important ingredient in most of these previous magnonic works is the presence of Dzyaloshinskii-Moriya interaction~\cite{Dzyaloshinsky1958}. However, the latter is only present in non-centrosymmetric systems and is related to the underlying spin-orbit coupling~\cite{PhysRevB.105.L180414} \--- which can be substantially small in real materials. Furthermore, specific details of the \textit{spin-magnon} mapping play an important role in finite-size bosonic systems, which is typically absent in the fermionic counterparts. Bosonic excitations carry a distinct feature \--- they must be positive definite. This unique characteristic results in specific onsite terms in the spin-magnon mapping. To the best of our knowledge, this fact has not been effectively considered in the previous studies~\cite{PhysRevX.8.011010} as far as higher-order topology is concerned. 
    
Due to the reasons outlined above, here, we focus on a centrosymmetric ferromagnetic (FM) system within a decorated 2D honeycomb lattice configuration, as shown in Fig.~\ref{fig:Fig1}. This setup can be envisioned as an extension of the one-dimensional Su-Schrieffer-Heeger model. Within this context, two distinct distortion scenarios naturally emerge \--- one where the inter-unit cell coupling surpasses the coupling within a unit cell ($J_0 < J_1$), resulting in a kekul\'e structure; the alternate scenario entails an anti-kekul\'e distortion, arising when the opposite limit is considered~\cite{Hatsugai2019}. In our study of the kekul\'e structure, we identify two distinct SOTMI phases. The first is a \textit{intrinsic} SOTMI phase, which features conventional protected corner modes. The second is a \textit{pseudo}-SOTMI phase that emerges when the system is truncated differently. Surprisingly, we also discover bond-localized magnon modes in the anti-kekul\'e structure. We analyze the effects of disorder to assess the robustness of these characteristic boundary states. Our findings indicate that the boundary modes in the kekul\'e configuration are resilient, while the bond-localized modes are not. 
      
%-------------------------------------------------------------------------------------------
\begin{figure}[t!]
\centering \includegraphics[width=\columnwidth]{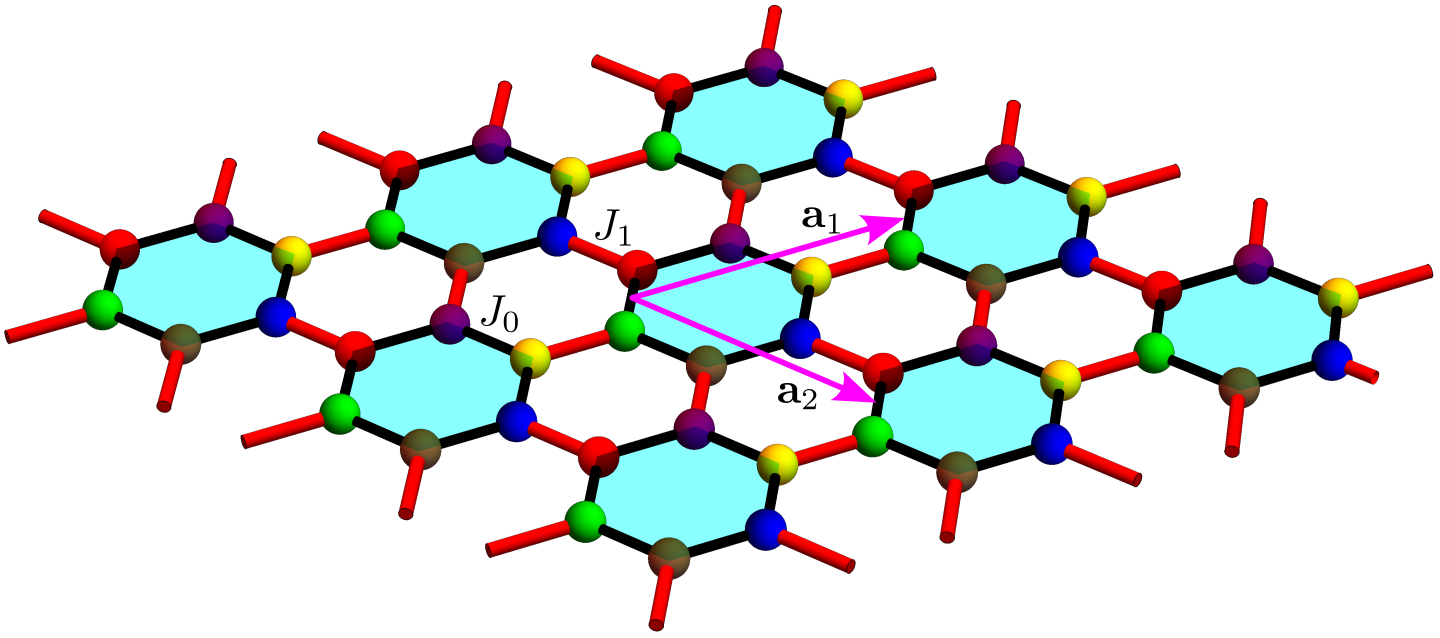}
\caption{Schematic representation of a 2D honeycomb-lattice model with a six-site unit cell (depicted by colored hexagons). The black bonds represent ferromagnetic coupling with strength $J_0$, while the red bonds indicate coupling with strength $J_1$. The lattice vectors ${\bf{a}}_1$ and ${\bf{a}}_2$ are shown, along with the sublattice sites (distinguished by six different colors) as labeled in the figure.}\label{fig:Fig1}
\end{figure}
%-------------------------------------------------------------------------------------------
    
{\textcolor{blue}{\textit{Model and method}}}\--- The system of our interest involves nearest neighbor Heisenberg spin exchange interactions on a decorated honeycomb lattice comprising of six sub-lattices with intra-cell (inter-cell) exchange couplings denoted by $ J_0~(J_1)$ (see Fig.~\ref{fig:Fig1}). The corresponding Hamiltonian is given by
%----------------------------
\begin{equation}\label{eq.1}
\mathcal{H} = - \sum_{\langle ij \rangle} J_{ij} {\bf{S}}_i \cdot {\bf{S}}_j - \beta \sum_{ i } (S^z_{i})^2 \ ,
\end{equation}
%----------------------------
where, $J_{ij}$'s are equal to $J_0$ or $J_1$ depending on the bond-types, and $\beta > 0$ signifies the strength of the onsite single-ion anisotropy. Note that, we account for the onsite anisotropy term as the Mermin-Wagner theorem prohibits long-range magnetic ordering in two dimensions. To obtain the magnon picture in the linear spin-wave regime, we carry out the spin-magnon mapping by performing Holstein-Primakoff transformation on the FM ground state, given by
%---------------------------------
\begin{equation}
\label{eq.2}
S^{+}_i \approx \sqrt{2S}  \eta_{i\alpha}, \; 
S^{-}_i \approx \sqrt{2S}\eta_{i\alpha}^{\dagger}, \;
S^{z}_i = \left( S - \eta_{i\alpha}^{\dagger} \eta_{i\alpha} \right).
\end{equation}
%---------------------------------
Here, $S^{\pm}_i = S^{x}_i \pm i S^{y}_i$, are the raising and lowering operators, and $\eta^\dagger_{i\alpha}$ creates a magnon excitation at site $i$. Since we have six sublattice sites, there are six flavors ($\alpha = 1 \ldots 6$) of magnon operators. The corresponding bosonic Hamiltonian reads as $\mathcal{H} = \mathcal{H}_0 + \mathcal{H}_1$ where
%---------------------------------------
\begin{subequations}
\begin{align}
\label{eq.3.1}
\mathcal{H}_0 & = 
S\sum_{\langle ij \rangle} J_{ij} (\eta^{\dagger}_i\eta_i+\eta^{\dagger}_j\eta_j)
+
2\beta S \sum_{i}\eta^{\dagger}_i\eta_i\ , \\
\label{eq.3.2}
\mathcal{H}_1 & = 
-S\sum_{\langle ij \rangle} J_{ij}(\eta^\dagger_i \eta_j + \rm{h.c.})\ ,
\end{align}
\end{subequations}
%---------------------------------------
Here, $\mathcal{H}_0$ corresponds to the onsite part, and $\mathcal{H}_1$ denotes the magnon hopping contribution to the total Hamiltonian. The dependence on the sublattice degrees of freedom is implicitly assumed. Furthermore, $J_{ij}>0$ is a necessary constraint needed to ensure FM order. The first term of the Hamiltonian in Eq.~\eqref{eq.3.1} indicates bond-dependent onsite energy highlighting an intrinsic property of \textit{magnonic} Hamiltonians in contrast to the usual \textit{fermionic} tight-binding Hamiltonians. \SW{The onsite energy depends on the number of neighboring bonds, and hence is different for the bulk and the boundary sites. This plays a crucial role in determining the nature of the boundary modes for finite systems, as we explain in more detail in the following text.}
    
%-------------------------------------------------------------------------------------------
\begin{figure*}[t!]
\centering \includegraphics[width=\linewidth]{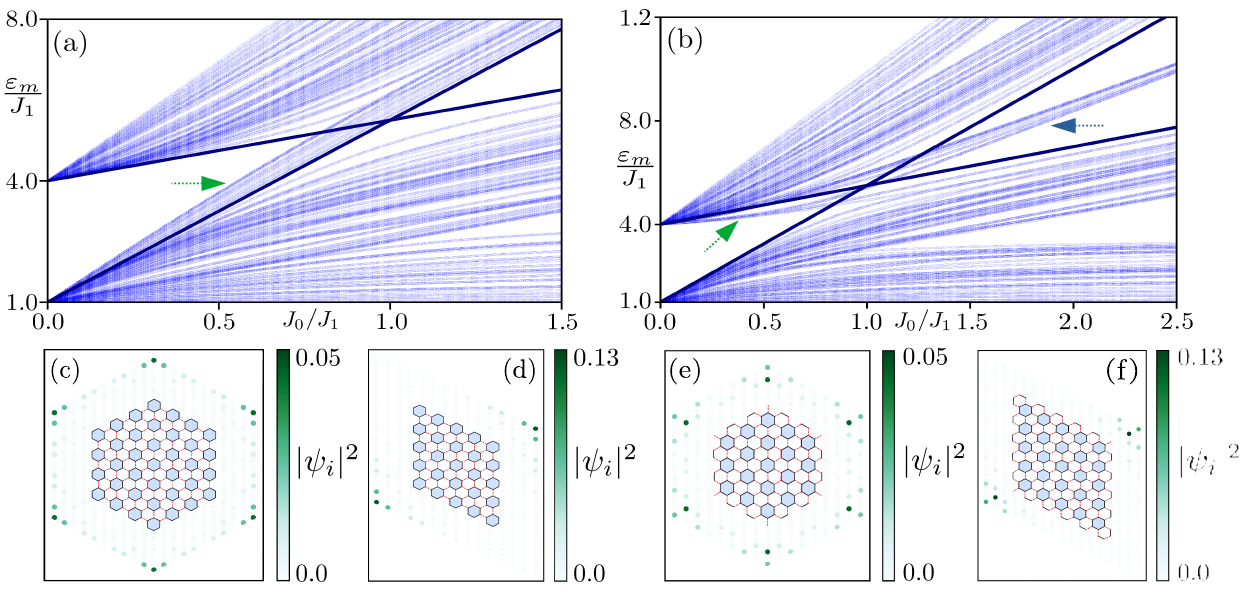}
\caption{Numerical results for a finite-size system are illustrated employing open boundary conditions (OBC). The panels (a) and (b) depict the magnon eigenvalue spectra as a function of $J_0/J_1$ for hexagonal geometries corresponding to the lattice structures depicted in the insets of panels (c) and (e), respectively. The dark blue lines in the spectra represent the two energy eigenvalues of $\mathcal{H}_\vk$ at $\vk=\Gamma$, identifying the magnon bulk gap $\Delta$. Panels (c) and (d) illustrate site-resolved normalized probabilities ($|\psi_i|^2$) of in-gap states contributing to MCMs for $J_0=0.7$, with the respective lattice geometries shown in the insets. $|\psi_i|^2$ is averaged over six (two) states contributing to MCMs for hexagonal (rhombus) geometry. Panels (e) and (f) depict the averaged site-resolved normalized probabilities of in-gap states representing the pseudo corner modes for $J_0=0.7$, with the corresponding lattice geometries displayed in the insets. The finite-size lattice geometries in (c) and (d) consist of 222 sites (37 unit cells) and 150 sites (25 unit cells), respectively, representing the intrinsic SOTMI phase. The lattice geometries in (e) and (f) consist of 144 sites (12 outer-edge bonds) and 192 sites (20 outer-edge bonds), respectively, demonstrating the pseudo SOTMI phase. Other model parameters are chosen as $(J_1, \beta, S)=(1, \tfrac{1}{3}, \tfrac{3}{2})$.} \label{fig:Fig2}
\end{figure*}
%-----------------------------------------------------
    
{\textcolor{blue}{\textit{Realization of SOTMI phase and its topological characterization}}} \--- Assuming translational invariance, the Hamiltonian in Eq.~\eqref{eq.3.1}-\eqref{eq.3.2} can be written in momentum space as $\mathcal{H}=\sum_\vk \psi^\dagger_\vk\mathcal{H}_\vk\psi_\vk$, where $\mathcal{H}_\vk$ is a 6$\times$6 matrix having basis $\psi_\vk=(\eta_{\vk,1},\eta_{\vk,2},\eta_{\vk,3},\eta_{\vk,4},\eta_{\vk,5},\eta_{\vk,6})^{\sf{T}}$~[see the supplementary material (SM)~\cite{supp} for details]. It is easy to check that $\mathcal{H}_\vk$ respects time-reversal symmetry: $ \mathcal{H}^{*}_\vk  = \mathcal{H}_{-\vk}$, inversion symmetry $\mathcal{P}=\sigma_x\otimes \mathcal{I}_3 $: $\mathcal{P}^{-1}\mathcal{H}_\vk\mathcal{P}=\mathcal{H}_{-\vk}$ along with a six-fold rotational (${\sf{C}}_6$) symmetry: $U^\dagger_{{\sf{C}}_6}\mathcal{H}_\vk U_{{\sf{C}}_6}=\mathcal{H}_{{\sf{C}}_6 \vk}$. Here, $U_{{\sf{C}}_6}$ corresponds to the unitary operator representation for ${\sf{C}}_6$ rotation and $\sigma_x,\mathcal{I}_3 $ denote Pauli matrix and Identity matrix respectively~\cite{supp}. \SW{Under these symmetry considerations, we adopt the well-developed mathematical framework of symmetry indicator topological invariant~\cite{PhysRevB.99.245151,Benalcazar2017,benalcazarprb2017,PhysRevX.8.031070} to characterize of higher-order topology of the magnonic system. Considering the $\sf{C}_6$ symmetry we label this invariant as $\chi^{(6)}$ (see the details in Ref.~\cite{supp}).}
    
The magnon band structure is obtained by diagonalizing $\mathcal{H}_\vk$. \SW{We notice that for $J_0 \neq J_1$ , the bulk spectrum is gapped at finite energy in the mid-band region (between the third and the fourth band)}~\cite{supp}. Note that, the gap opens at $\Gamma$ point in contrast to the conventional Dirac point for graphene. In this case, we obtain an analytical expression for the associated bulk gap as : $\Delta=2S\lvert J_0-J_1\rvert$. Consequently, $\mathcal{H}_\vk$ goes through a gap-closing transition at $\vk=\Gamma$ for $J_0=J_1$~\cite{supp} which is at the middle of the magnon bandwidth. \SW{Furthermore, when $J_0 > 1.32 J_1 $, a secondary  trivial bulk gap appears away from the mid band region both at low energy (between the first and the second band) and high energy (between the fifth and the sixth band).}
    
To appropriately distinguish the SOTMI phase from the trivial gapped phase, we employ the symmetry indicator integer topological invariant $\mathcal{\chi}^{(6)}$ for proper characterization of the magnon Bloch bands~\cite{PhysRevB.99.245151}. Here, $\mathcal{\chi}^{(6)}$ is constructed by utilizing the previously mentioned symmetries of $\mathcal{H}_\vk$ as well as the symmetries of the high-symmetry points in the Brillouin zone [see the SM~\cite{supp} for details]. $\mathcal{\chi}^{(6)}\not=0$ identifies the SOTMI phase that ensures the presence of topologically protected magnon corner modes (MCMs), while the trivial phase is marked by $\mathcal{\chi}^{(6)}=0$. Our system displays the SOTMI phase hosting topologically protected MCMs in the regime $J_0 < J_1$ (kekul\'e structure) with $\mathcal{\chi}^{(6)}=2$. \SW{In addition, we rule out the signatures of any first-order topological phase by computing the appropriate topological invariant ~\cite{benalcazarprb2017,Benalcazar2017,vanderbilt_2018}. (see SM~\cite{supp} for the details).}
    
{\textcolor{blue}{\textit{Emergence of topologically protected MCMs in finite-size systems}}}\--- We capture the topological characteristics of the SOMTI phase hosting the MCMs by applying open boundary conditions (OBCs) and performing numerical calculations based on the finite-size 2D lattice geometry. In this context, two distinct lattice geometries are considered: \textit{hexagonal}, and \textit{rhombus} having six and two corners, respectively (see the insets in Fig.~\ref{fig:Fig2}). A few remarks are necessary at this stage: (i) As a result of bond-dependent onsite energy, the first term in Eq.~\eqref{eq.3.1}  readily generates an onsite energy difference (OED) between the bulk and the boundary by an amount $SJ_1$, and (ii) the OED reflects how the edges are terminated in different geometries. Such a feature is generally absent in a similar fermionic system.
    
The energy eigenvalue spectrum as a function of $J_0/J_1$ is obtained by diagonalizing the Hamiltonian in Eq.~\eqref{eq.3.1}-\eqref{eq.3.2} with OBC [see Fig.~\ref{fig:Fig2}(a)]. \SW{Note that the spectrum originates from the finite value because of a finite onsite anisotropy term with finite $\beta$. In the region $J_0<J_1$ (kekul\'e structure), there exists identifiable in-gap states  demonstrating topologically protected MCMs, whereas there exists no in-gap states when $J_0>J_1$ (anti-kekul\'e structure). This reflects consistency with the obtained value of $\chi^{(6)}$.} Additionally, the in-gap states are closer to the lower bulk states for $J_0 \ll J_1$ while they are closer to both the upper and lower bulk states when $J_0 \lesssim J_1$. The states immediately below the upper bulk states [indicated by the arrow in Fig.~\ref{fig:Fig2}(a)] contribute to the MCMs. This reflects the consequence of OED and reveals a true \textit{magnonic} signature distinguishing it from the regular \textit{fermionic} systems, where one would readily expect the topologically protected in-gap states to appear precisely in the middle of the bulk gap. Specifically, we obtain six (two) states that exhibit MCM signatures while considering hexagonal (rhombus) lattice geometry. We depict the localized MCMs by the normalized site-resolved probability $\lvert \psi_i\rvert^{2}$ of the particular in-gap states in Fig.~\ref{fig:Fig2}(c,d), with respective lattice geometry illustrated in the inset. \SW{The other in-gap states do not contribute to the MCMs and are not zero-dimensional. A few of them are shown in SM~\cite{supp}, and resembles fractal like structures. However, we skip the discussion on these states as it is beyond the scope of the present paper.}
    
Thus far, our finite-size analysis has followed a specific termination scheme, ensuring that individual unit cells remain intact (type I), as demonstrated in the insets of Fig.~\ref{fig:Fig2}(c,d). This arrangement leads to what we call the \textit{intrinsic} SOTMI phase, for which the results of our numerical analysis have been displayed in Fig.~\ref{fig:Fig2}(a,c,d). However, another scenario, type II, naturally emerges when the sample edges are terminated differently, resulting in incomplete unit cells. In this configuration, there are six (two) sites at the corners with single bonds for the hexagonal (rhombus) geometry, as depicted in the insets of Fig.~\ref{fig:Fig2}(e,f). This arrangement gives rise to a \textit{pseudo}-SOTMI phase, originating from the in-gap states present in the topological region ($J_0 < J_1$), as indicated by the green arrow in Fig.~\ref{fig:Fig2}(b). Analyzing the site-resolved normalized probability of these states for both geometries, it becomes clear that these states do not exhibit robust localization exclusively at the corners [see Fig.~\ref{fig:Fig2}(e,f)]. Instead, they display partial localization, with the highest probability contribution occurring at the sites coupled to the corners. This observation justifies the term \textit{pseudo}-SOTMI phase. As explained earlier, this analysis differs significantly from previously reported theoretical work on an analogous fermionic system~\cite{Hatsugai2019}.
    
%-------------------------------------------------------------------------------
\begin{figure}[t!]
\centering \includegraphics[width=\columnwidth]{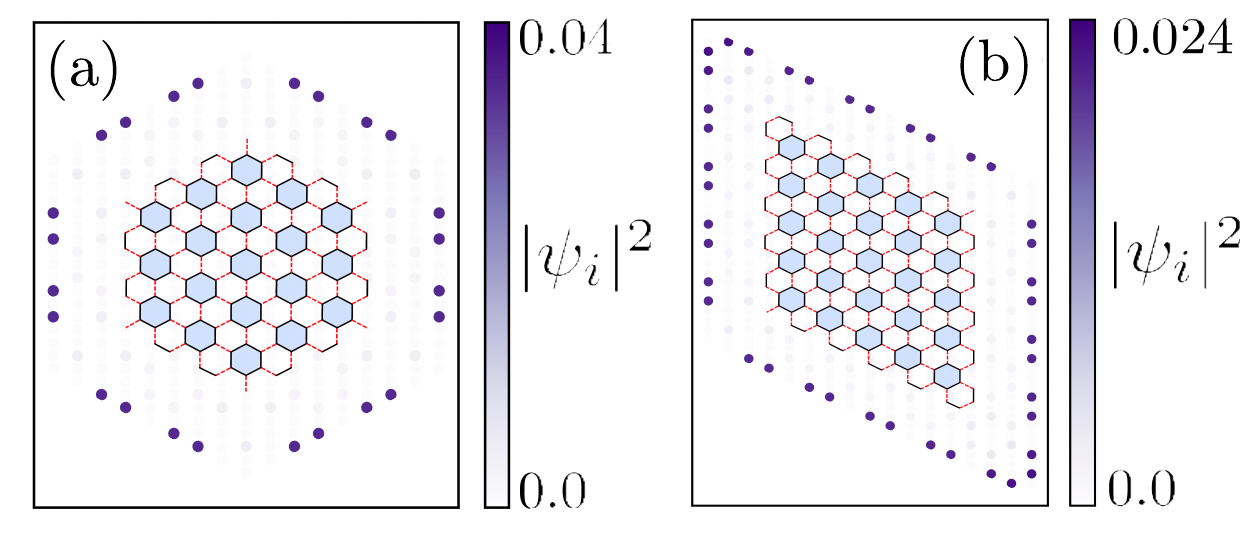}
\caption{Site-resolved normalized probability distributions of in-gap states illustrating \SW{\textit{Tamm/Shockley}} type BLMMs for $J_0=2.0$, shown in panels (a) and (b) for hexagonal and rhombus geometries respectively. The lattice geometries in (a) and (b) consist of 144 sites (12 outer-edge bonds) and 192 sites (20 outer-edge bonds), respectively. Other model parameters are chosen as $(J_1, \beta, S)=(1, \tfrac{1}{3}, \tfrac{3}{2})$.}\label{fig:Fig3}
\end{figure}
%-------------------------------------------------------------------------------
Further analyzing the magnon spectrum in the type II scenario and as a function of $J_0/J_1$ in the non-topological regime ($J_0 > J_1$) [see Fig.~\ref{fig:Fig2}(b)], we note that there also exist in-gap states. The number of these in-gap states depends on the system size and geometry, precisely matching the number of bonds constituting the outer edge of the finite system. They lead to interesting geometric  \SW{\textit{Tamm/Shockley}} type bond-localized magnon modes (BLMMs). These BLMMs can be visualized through the $\lvert \psi_i \rvert^{2}$ of the in-gap states depicted in Fig.~\ref{fig:Fig3}(a,b), corresponding to the respective lattice geometry shown in the inset. However, they lack any topological protection as $\mathcal{\chi}^{(6)} = 0$ in this case. 

\SW{Note that the apparent asymmetry of the in-gap states in Fig.~\ref{fig:Fig2}(a,b) is a direct consequence of the OED as mentioned earlier. This is in stark contrast to its fermionic analogs where typically the in-gap states are positioned at the middle of bulk gaps for particle-hole symmetric systems.}
    
%------------------------------------------------------------------------------------------------
\begin{figure}[t!]
\centering \includegraphics[width=\columnwidth]{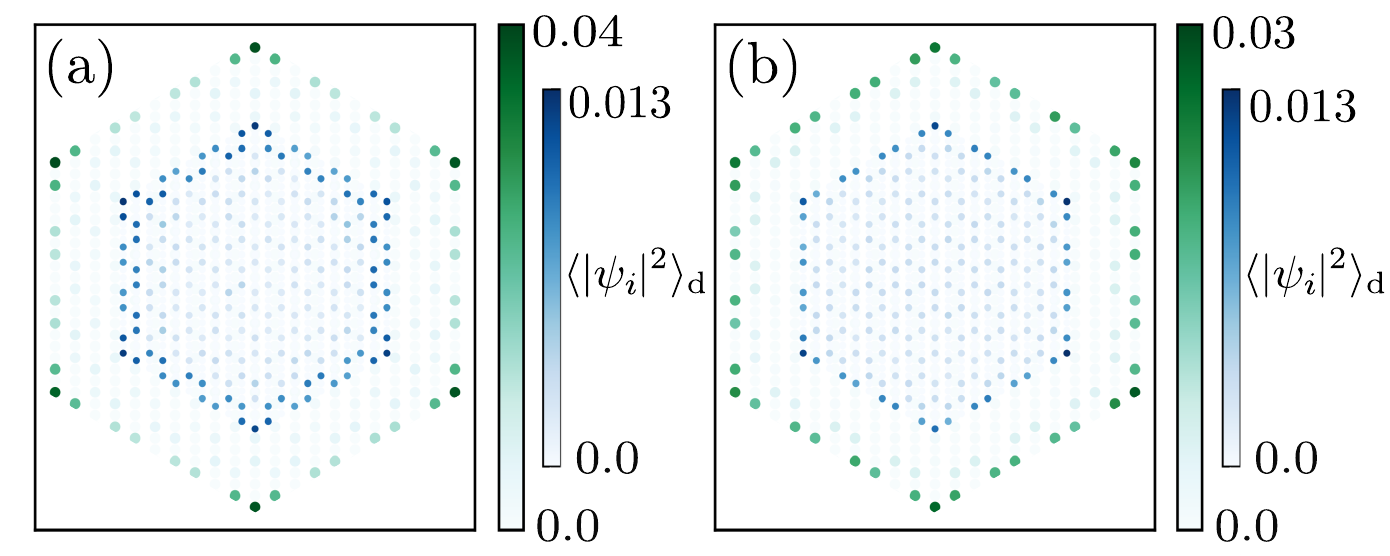}
\caption{Disorder-averaged site-resolved probability distribution $\langle|\psi_i|^2\rangle_{\rm{d}}$ of states leading to MCMs is depicted in the intrinsic SOTMI phase. (a) Depicts Case I for $\zeta=0.8$ and $J_0=0.7$, with the inset corresponding to $\zeta=0.8$ and $J_0=0.4$. (b) Illustrates Case II for $\zeta=0.3$ and $J_0=0.3$, with the inset displaying for $\zeta=0.7$ and $J_0=0.7$. We choose the other model parameter values as $(J_1, \beta, S)=(1, \tfrac{1}{3}, \tfrac{3}{2})$.}
\label{fig:Fig4}
\end{figure}
%------------------------------------------------------------------------------------------------

{\textcolor{blue}{\textit{Effect of disorder and stability analysis for MCMs}}}\--- Up to this point, we discuss both the topological and non-topological characteristics of the honeycomb network by considering clean (disorder-free) limits. However, real materials are prone to impurities and disorder. Consequently, without aiming for a specific material realization, we focus on a generic disorder by considering a random out-of-plane exchange anisotropy. The Hamiltonian in Eq.~\eqref{eq.1} is therefore modified as $\mathcal{H} \Rightarrow \mathcal{H} + \mathcal{H}_{\rm{dis}}$, where 
%---------------------------
\begin{equation}\label{eq.4}
\mathcal{H}_{\rm{dis}} = - \sum_{\langle ij \rangle} J^{\rm{d}}_{ij} S^z_i S^z_j\ .
\end{equation}
%---------------------------
Here, $J^{\rm{d}}_{ij}$ is a random coupling parameter uniformly distributed in the range $\left[-\zeta,\zeta\right]$ with $\zeta$ being the disorder strength. Ensuring FM order in the presence of disorder readily imposes the following constraint:  $\zeta_{max} = J_0 (J_1)$ for intra (inter) cell exchange coupling. Remarkably, the out-of-plane exchange disorder mentioned in Eq.~\eqref{eq.4} leads to an effective bond-dependent on-site disorder for the bosonic case. This translates into a disordered magnon Hamiltonian where we substitute $J_{ij}$ by $J_{ij} + J^{\rm{d}}_{ij}$ in Eq.~\eqref{eq.3.1}.
    
In the subsequent discussions, we focus primarily on the {\it{intrinsic}} SOTMI phase by investigating the stability of the MCMs (magnon corner modes) in the presence of such a disorder, as mentioned earlier. The qualitative features of the {\it{pseudo}} SOTMI phase remain identical to the previous case. We analyze the disorder-averaged site-resolved probability $\braket{| \psi_i|^2}_{\rm{d}}$ for specific states while varying the disorder strength $\zeta$. We investigate two cases with distinct characteristics:
    
Case I: Disorder in the inter-cell coupling $J_1$: In this scenario, the system continues to exhibit MCM signatures for higher values of $J_0$ even for higher disorder strength $\zeta$ [see Fig.~\ref{fig:Fig4}(a)].
    
Case II: Disorder in the intra-cell coupling $J_0$: Here, the system tends to exhibit outer-edge bond localization when the disorder is introduced [see Fig.~\ref{fig:Fig4}(b)].
    
For both cases, the MCMs are more susceptible to the disorder effects for smaller $J_0$ than a larger value with the same disorder strength. This observation arises from the fact that the in-gap states, responsible for the emergence of MCMs, at lower $J_0$ values are closer to the bulk states [see Fig.~\ref{fig:Fig2}(a)]. Consequently, these MCMs tend to hybridize with the bulk states even in the presence of relatively mild disorder and is eventually destroyed. We refer the readers to the SM~\cite{supp} for the discussion of stability analysis of MCMs in the case of rhombus geometry. 
    
{\textcolor{blue}{\textit{Summary and Discussion}}} \--- To summarize, in this Letter, we introduce a FM honeycomb lattice network that enables the realization of a higher-order bosonic topological system. Our findings are particularly relevant to centrosymmetric systems and underscore the crucial role of spin-magnon mapping in generating both an intrinsic and a pseudo-SOTMI phase, each characterized by appropriate topological invariant and unique boundary modes. In this context, our work stands in stark contrast to a previous theoretical study~\cite{Pantaleon_2019}, which explores a similar system under non-centrosymmetric conditions. While intrinsic SOTMI phases have been previously discussed in magnonic systems (square lattice) in the presence of anti-skyrmions~\cite{PhysRevLett.125.207204, PhysRevB.104.024406}, our study exhibiting the emergence of both intrinsic and pseudo-SOTMI phase in a simple magnonic model system carries significant contribution, to the best of our knowledge.
    
Moreover, we have explored the robustness of the boundary modes within the topological phase under the influence of random disorder. Specifically, we model the disorder arising from out-of-plane exchange anisotropy and provide qualitative estimates of the topological robustness of MCMs across varying disorder strengths. It is worth noting that during the preparation of this manuscript, we became aware of a recent theoretical work~\cite{PhysRevB.105.L180414}, where the authors investigated a different disorder realization involving onsite magnetic fields. Our work complements this research by offering insights into a more realistic scenario through an exchange anisotropy disorder. From a practical point of view, the spatial distribution of the localized MCMs can possibly be measured by nitrogen-vacancy center magnetometry~\cite{10.1063/1.5141921} or near-field Brillouin light scattering~\cite{10.1063/1.3502599}. Overall, our findings hold promise for advancing robust future magnonic devices.
       
{\textcolor{blue}{\textit{Acknowledgments}}} \--- We thank Jason T. Haraldsen and Ying Su for providing important feedback while preparing the manuscript. Sayak Bhowmik (S.B.) and A.S. acknowledge the SAMKHYA: HPC Facility provided at IOP, Bhubaneswar, for numerical computations. Saikat Banerjee (S.B.) acknowledges support from the U.S. Department of Energy (DOE), Office of Science, and Office of Advanced Scientific Computing Research through the Quantum Internet to Accelerate Scientific Discovery Program, and partial support from the Office of Basic Energy Sciences, Material Sciences and Engineering Division, U.S. DOE under Contract No. DE-FG02-99ER45790.

%%%%%%%%%%%%%%%%%%%%%%%%%%%%%%%%%%%%%%%%%%%%%%%%%%%%%%%%%%%%%%%%%%%%%%%%%%%%%
%\bibliographystyle{apsrev4-1}
\bibliography{ref}{}
%%%%%%%%%%%%%%%%%%%%%%%%%%%%%%%%%%%%%%%%%%%%%%%%%%%%%%%%%%%%%%%%%%%%%%%%%%%%%

\clearpage

\newpage

\begin{onecolumngrid}
\begin{center}
{
\fontsize{12}{12}
\selectfont
\textbf{Supplemental material for ``Higher-order topological corner and bond-localized modes in magnonic insulators''\\[5mm]}
}
\normalsize Sayak Bhowmik$^{1,2}$, Saikat Banerjee$^{3,4}$, and Arijit Saha$^{1,2}$\\
\vspace{2mm}
{\small $^1$\textit{Institute of Physics, Sachivalaya Marg, Bhubaneswar-751005, India}\\[0.5mm]}
{\small $^2$\textit{Homi Bhabha National Institute, Training School Complex, Anushakti Nagar, Mumbai 400094, India}\\[0.5mm]}
{\small $^3$\textit{Theoretical Division, T-4, Los Alamos National Laboratory, Los Alamos, New Mexico 87545, USA}\\[0.5mm]}
{\small $^4$\textit{Center for Materials Theory, Rutgers University, Piscataway, New Jersey, 08854, USA}\\[0.5mm]}

\end{center}

%=========================================================================================================================================================================================
\renewcommand{\thesection}{S\arabic{section}}
%\maketitle
\tableofcontents 

%---------------------------------------------------------------
\section{Linear spin-wave theory and magnons \label{sec:sec_1}}
%---------------------------------------------------------------
In this section, we provide the details of the linear spin-wave theory to analyze the magnonic band structure arising from the decorated honeycomb lattice system introduced in the main text. 
Consequently, following a Holstein-Primakoff transformation (see main text), the corresponding Hamiltonian can be written in terms of the magnon (bosonic) creation and annihilation operators as 
$\mathcal{H} = \mathcal{H}_0 + \mathcal{H}_1$, where

%---------------------
\begin{subequations}
\begin{align}
\label{eq.1.1}
\mathcal{H}_0 & =
S\sum_{\langle ij \rangle} J_{ij} (\eta^{\dagger}_i\eta_i+\eta^{\dagger}_j\eta_j)
+
2\beta S \sum_{i}\eta^{\dagger}_i\eta_i\ , \\
\label{eq.1.2}
\mathcal{H}_1 & = 
-S\sum_{\langle ij \rangle} J_{ij}(\eta^\dagger_i \eta_j + \rm{h.c.})\ . 
\end{align}
\end{subequations}
%---------------------
Note that $\mathcal{H}_0$ signifies the onsite part, while $\mathcal{H}_1$ denotes the corresponding magnon hopping within the lattice. It can be readily written in momentum space as $\mathcal{H}=\sum_\vk \psi^\dagger_\vk\mathcal{H}_\vk\psi_\vk$, where  $\psi_\vk=(\eta_{\vk,1},\eta_{\vk,2},\eta_{\vk,3},\eta_{\vk,4},\eta_{\vk,5},\eta_{\vk,6})^{\sf{T}}$. The Bloch-form Hamiltonian $\mathcal{H}_\vk$ is written as follows
%-----------------------------------
\begin{equation}\label{eq.2}
\mathcal{H}_{\vk} 
= S
\begin{pmatrix}
\epsilon_0 & -J_0 & 0 & -J_1 e^{i\vk \cdot (\va_1-\va_2)} & 0 & -J_0 		\\
-J_0 & \epsilon_0 & -J_0 & 0 & -J_1 e^{i\vk \cdot \va_1} & 0				\\
0 & -J_0 & \epsilon_0 & -J_0 & 0 & -J_1 e^{i\vk \cdot \va_2}				\\
-J_1 e^{-i\vk \cdot (\va_1-\va_2)} & 0 & -J_0 & \epsilon_0 & -J_0 & 0 		\\   
0 & -J_1 e^{-i\vk \cdot \va_1} & 0 & -J_0 & \epsilon_0 & -J_0        	\\
J_0 & 0 & -J_1 e^{-i\vk \cdot \va_2} & 0 & -J_0 & \epsilon_0                     
\end{pmatrix}\ ,
\end{equation}
%-----------------------------------
where $\epsilon_0=(2J_0+J_1)+2\beta$ and $\va_1, \va_2$ are the two lattice vectors as defined in Fig.~1 in the main text. The magnon spectrum can be obtained by diagonalizing the above Hamiltonian. Such a magnonic Bloch band structure is shown along with the gap closing transition (at $\Gamma$ point in the Brillouin zone) of $\mathcal{H}_{\vk}$ in Fig.~\ref{fig:Fig1}. Note that, the above Hamiltonian respects six-fold rotational symmetry as $U^{\dag}_{{\sf{C}}_6} \mathcal{H}_\vk U_{{\sf{C}}_6} = \mathcal{H}_{{\sf{C}}_6 \vk}$. The unitary operator representation of this six-fold rotational symmetry, $U_{{\sf{C}}_6}$, reads as
%------------------------------
\begin{equation}\label{eq.3}
U_{{\sf{C}}_6}
=
\begin{pmatrix}
0 & 0 & 0 & 0 & 0 & 1	\\
1 & 0 & 0 & 0 & 0 & 0	\\
0 & 1 & 0 & 0 & 0 & 0	\\
0 & 0 & 1 & 0 & 0 & 0	\\
0 & 0 & 0 & 1 & 0 & 0	\\
0 & 0 & 0 & 0 & 1 & 0
\end{pmatrix}\ .
\end{equation}
%------------------------------

%-----------------------------------------------------------------------
\begin{figure}[t!]
\centering \includegraphics[width=1.0\linewidth] {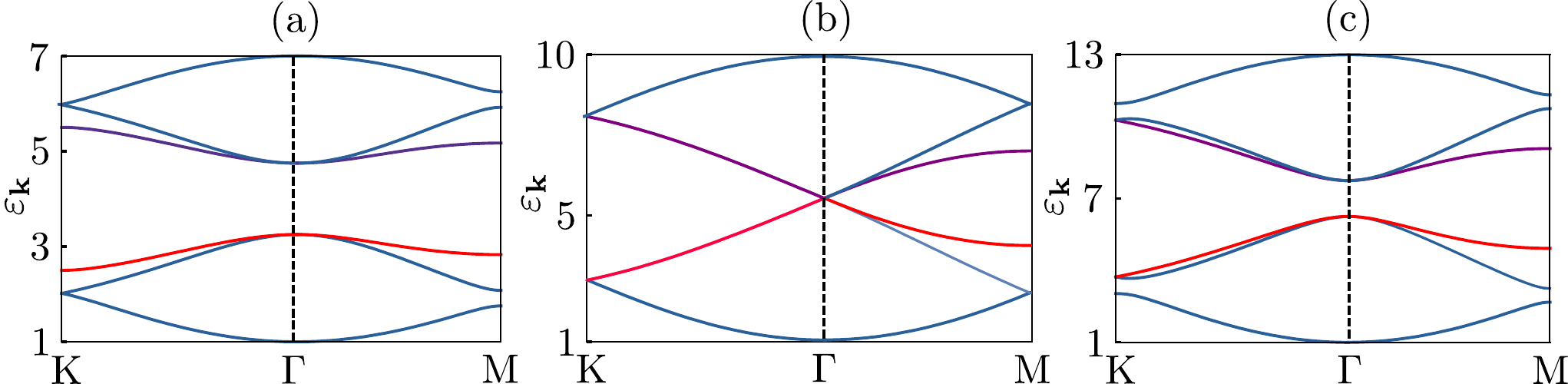}
\caption{The magnon band spectrum along the line joining high-symmetry points ($\rm{K}, \Gamma,\rm{ M}$) in the Brillouin zone is depicted for $(J_0, J_1)$= (a) ($0.5, 1$), (b) ($1, 1$), (c) ($1.5, 1$).  Other model parameters are chosen as $(\beta, S)=( \tfrac{1}{3}, \tfrac{3}{2})$.  }\label{fig:Fig1}
\end{figure}
%-----------------------------------------------------------------------

%------------------------------------------------------------------------------------------------
\section{Symmetry indicator: Analysis of higher-order topological invariant \label{sec:sec_2}}
%------------------------------------------------------------------------------------------------
Exploiting the symmetries associated with the bulk Hamiltonian and the rotational symmetries of different high-symmetry points (HSP) in the Brillouin zone, we can construct the symmetry indicator topological invariant~\cite{PhysRevX.8.031070,PhysRevB.99.245151}. Given that an $n$ fold symmetric HSP $\Pi^{(n)}$ in the Brillouin zone satisfies the commutation relation: $[U_{{\sf{C}}_n},\mathcal{H}_{\Pi^{(n)}}]=0$, where $U_{{\sf{C}}_n}$ is the unitary representation of $n$ fold symmetry with eigenvalues $U^{p}_{{\sf{C}}_n}=e^{2\pi i(p-1)/n}$, ($p=1,2,..,n$), one can construct the rotation matrix at $\Pi^{(n)}$ for a given set of occupied Bloch states as
%------------------------------
\begin{equation}\label{eq.4}
\mathcal{S}^{ij}_{\Pi_{(n)}}=\bra{u_{i}(\Pi^{(n)})}U_{{\sf{C}}_n} \ket{u_{j}(\Pi^{(n)})}\ ,    
\end{equation}
%------------------------------
where $\ket{u_{i}(\Pi^{(n)})}$ represents the Bloch states at $\vk=\Pi^{(n)}$ and $i,j$ runs over the occupied subspace of $\mathcal{H}_k$. By definition, the presence of non-trivial topology is identified when the rotational eigenvalues at $\Pi^{(n)}$ differ from the rotational eigenvalues at the reference point $\Gamma^{(n)}$. Incorporating all these, the integer topological invariant is defined as
%------------------------------
\begin{equation}\label{eq.5}
[\Pi_p^{(n)}] \equiv \lvert\#  \Pi_p^{(n)}-\# \Gamma_p^{(n)}\rvert\ ,    
\end{equation}
%------------------------------ 
where $\# \Pi_p^{(n)}$ reads as the number of eigenvalues of $\mathcal{S}_{\Pi_{(n)}}$ that are equal to $U^{p}_{{\sf{C}}_n}$. Invoking time-reversal symmetry and inversion symmetry constraints with 
the $C_6$ crystalline symmetry, the symmetry indicator topological invariant for the system of our interest is given as $\chi^{6}=([M_1^{(2)}],[K_1^{(3)}])$. However, we obtain $[K_1^{(3)}]=0$ for both 
the regions $(J_0<J_{1},J_0>J_1)$ for our system, hence we drop off the $[K_1^{(3)}]$ index in the main text for simplicity and retain  $[\ M_1^{(2)}]$ as the topological invariant.  

\vspace{0.3cm}
%------------------------------------------------------------------------
\begin{figure}[h!]
\centering \includegraphics[width=0.85\linewidth] {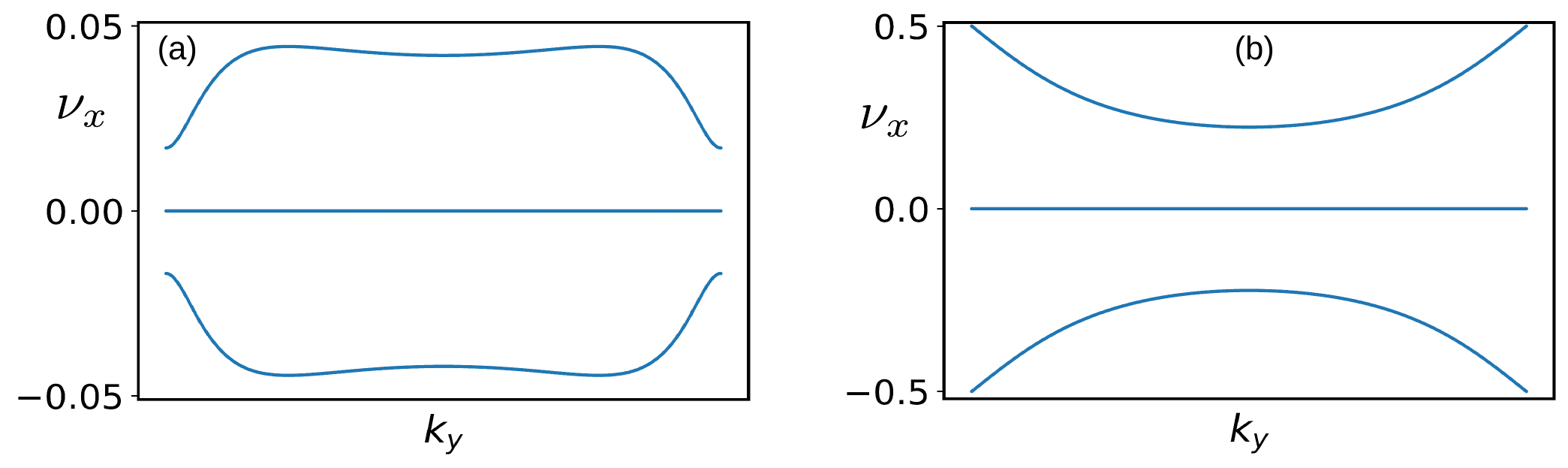}
\caption{The Wannier spectra is depicted as a function of $k_y$ choosing (a) $J_0=1.5$, and (b) $J_0=0.5$. The other model parameter values are chosen as  $(\beta, S, J_1)=( \tfrac{1}{3}, \tfrac{3}{2}, 1)$.  }
\label{fig:Fig2}
\end{figure}
%----------------------------------------------------------------------------

%----------------------------------------------------------------------------------------
\section{Wilson loop: Discussion on first-order topological invariant  \label{sec:sec_3}}
%----------------------------------------------------------------------------------------
Furthermore, we implement the Wilson loop formulation to compute the first-order topological invariant (the bulk-polarization)~\cite{PhysRevB.96.245115}. By utilizing the Bloch states of the occupied 
subspace of $\mathcal{H}_\vk$ within the Brillouin zone,  we can construct the Wilson loop operator ~\cite{PhysRevB.96.245115}  in the following way:
\begin{equation}\label{eq.6}
    \mathcal{W}_{k_x} = F_{k_x+(N_x-1)\Delta k_x, k_y }\cdots  F_{k_x+\Delta k_x, k_y } F_{k_x, k_y },
\end{equation}
where $[ F_{k_x, k_y }]_{ij}= \braket{\psi_{j,k_x+\Delta k_x, k_y} |\psi_{i,k_x, k_y}} $ with $\ket{\psi_{i,k_x, k_y}}$ representing  the $i$th occupied Bloch state of  $\mathcal{H}_\vk$ and  $\Delta k_x$ denotes the discretization of the Brillouin zone with $N_x$ number of discrete points along $k_x$. We obtain the Wannier spectra $2\pi\nu_x$ as a function of $k_y$ (see Fig.~\ref{fig:Fig2}) by diagonalizing the Wannier Hamiltonian: $\mathcal{H}_{\mathcal{W}_{k_x}}= -i$ln$\mathcal{W}_{k_x}$. Then, the bulk-polarization $p_{x}$ is given by~\cite{PhysRevB.96.245115}
%------------------------------
\begin{equation}\label{eq.7}
    p_x= \frac{1}{N_y}\sum_{k_y}\sum_{j} \nu^{j}_{x}(k_y)\ ,
\end{equation}
%------------------------------
where $N_y$ is the number of points along $k_y$ in the discretized Brillouin zone, and $j$ denotes the index for the occupied subspace. It is evident from Fig.~\ref{fig:Fig2} that  the Wannier spectra is symmetric around zero, implying  $p_x=0$ for both $(J_0<J_1, J_0>J_1)$. By following the similar procedure, we obtain $p_y=0$. This analysis implies the absence of any first-order topological phase in the system. 

%-----------------------------------------------------------------------------------------
\section{Stability analysis of magnon corner modes for rhombus geometry\label{sec:sec_4}}
%-----------------------------------------------------------------------------------------
In the main text, we mainly illustrate the disorder effects and stability analysis of magnon corner modes (MCMs) in the intrinsic second-order topological magnon insulator (SOTMI) phase for the hexagonal lattice geometry. In this section, we present a similar analysis for the rhombus geometry, and identical features are noted for this case as depicted in Fig.~\ref{fig:Fig3}.

%-----------------------------------------------------------------------------------------
\begin{figure}[t!]
\centering \includegraphics[width=0.5\linewidth] {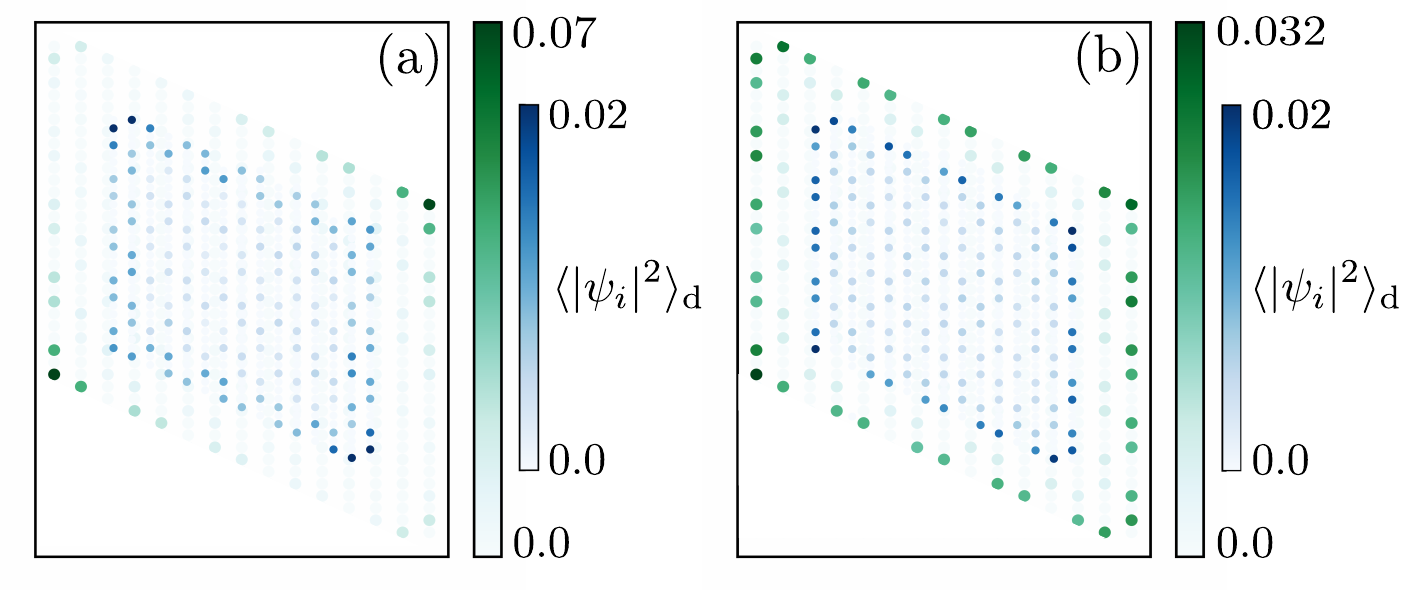}
\caption{Disorder-averaged site-resolved probability distribution ($\langle|\psi_i|^2\rangle_{\rm{d}}$) of states corresponding to MCMs is displayed in the intrinsic SOTMI phase. (a) Depicts Case I 
(intra-cell out of plane exchange disorder) for $\zeta=0.8$ and $J_0=0.7$, with the inset corresponding to $\zeta=0.8$ and $J_0=0.4$. (b) Illustrates Case II (inter-cell out of plane exchange disorder) 
for $\zeta=0.3$ and $J_0=0.3$, with the inset displaying for $\zeta=0.7$ and $J_0=0.7$. All other model parameters are chosen as $(J_1, \beta, S)=(1, \tfrac{1}{3}, \tfrac{3}{2})$.}
\label{fig:Fig3}
\end{figure}
%-----------------------------------------------------------------------------------------

%----------------------------------------------------------------------------------------
\section{Nature of the bulk states and the other in-gap states  \label{sec:sec_5}}
%----------------------------------------------------------------------------------------

%-----------------------------------------------------------------------------------------
\begin{figure}[h!]
\centering \includegraphics[width=1.0\linewidth] {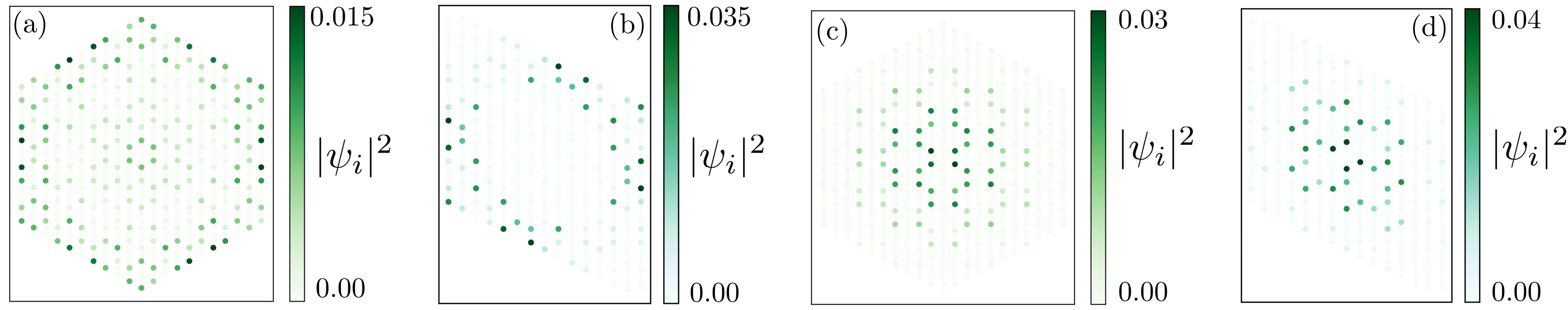}
\caption{Site-resolved normalized probability distributions ($|\psi_i|^2$) illustrating bulk states  for $J_0=2.0$ (non-topological regime) is shown in panels (a,c) and (b,d) for hexagonal and rhombus geometries respectively considering type-I termination of edge. Panels a,b represents $|\psi_i|^2$ of those states contributing to MCMs in the intrinsic-SOTMI phase. These states clearly represents bulk signatures when $J_0>J_1$. Panels c,d demonstrates the $|\psi_i|^2$ of the lower bulk states.  The lattice geometries in (a,c) and (b,d) consist of 222 sites (37 unit cells) and 150 sites (25 unit cells), respectively with type-1 termination. Other model parameters are chosen as $(J_1, \beta, S)=(1, \tfrac{1}{3}, \tfrac{3}{2})$.}
\label{fig:Fig4}
\end{figure}
%-----------------------------------------------------------------------------------------

In our study, we have identified three distinct boundary mode signatures: the \textit{intrinsic}-SOTMI, \textit{pseudo}-SOTMI, and the outer edge \textit{Tamm/Shockley} type bond-localized magnon modes (BLMMs). These unique modes manifest for specific terminations of the finite-size lattice, and the corresponding in-gap states contributing to these modes have been elaborately discussed in the main text. In this section we aim to depict the nature of the bulk states and the other in-gap states (the in-gap states other than those contributing the magnon corner modes (MCMs) in the \textit{intrinsic}-SOTMI phase). While considering the finite size system with type-I termination, it is clearly observed that there are no in-gap states in the non-topological regime $J_0>J_1$ [see Fig.~2(a) of main text]. Consequently, all the states for $J_0>J_1$ exhibits bulk like signatures as visually depicted by the $\langle|\psi_i|^2\rangle_{\rm{d}}$ of states in Fig.~\ref{fig:Fig4}.  Clearly, the states exhibiting MCM signatures in the intrinsic-SOTMI phase ($J_0<J_1$) with $\chi^6=2$, now represent bulk states in the regime $J_0 > J_1$ with $\chi^6=0$ as depicted in panel (a,b) of Fig.~\ref{fig:Fig4}. This  Additionally, the other low energy bulk states are shown in Fig.~\ref{fig:Fig4}(c,d).

Next, we consider the finite size system with termination type-II. We demonstrate $|\psi_i|^2$ for one of these bulk states in Fig.~\ref{fig:Fig5}. The other bulk states are visually similar to the ones shown in Fig.~\ref{fig:Fig5}. They do not demonstrate any localization property at the boundary.

The other in-gap states \--- apart from those exhibiting MCMs (\textit{intrinsic}-SOTMI phase)\--- are visually depicted in Fig.~\ref{fig:Fig6} by considering type-I edge-termination of the finite system.  These states demonstrates fractal-like structure, as shown in Fig.~\ref{fig:Fig6}. However, a comprehensive description and analysis of these fractal-like states are beyond the scope of our manuscript.

%-----------------------------------------------------------------------------------------
\begin{figure}[h!]
\centering \includegraphics[width=0.7\linewidth] {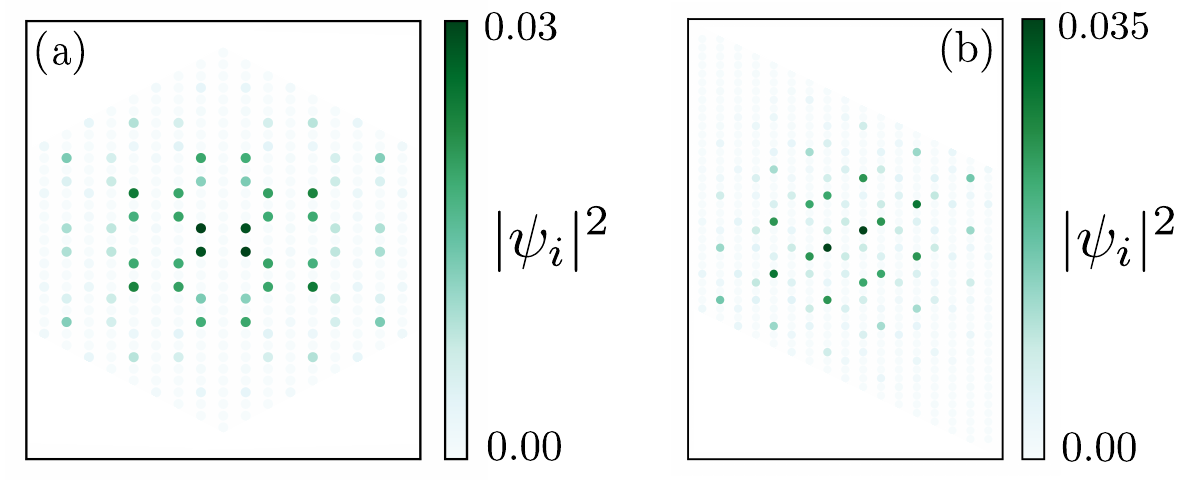}
\caption{Site-resolved normalized probability distributions ($|\psi_i|^2$) illustrating bulk states  for $J_0=2.0$ (non-topological regime) is shown in panels a and b  for hexagonal and rhombus geometries respectively considering type-II termination of edge. The lattice geometries in (a) and (b) consist of 144 sites (12 outer-edge bonds) and 192 sites (20 outer-edge bonds), respectively. Other model parameters are chosen as $(J_1, \beta, S)=(1, \tfrac{1}{3}, \tfrac{3}{2})$. }
\label{fig:Fig5}
\end{figure}
%-----------------------------------------------------------------------------------------

%-----------------------------------------------------------------------------------------
\begin{figure}[h!]
\includegraphics[width=1.0\columnwidth]{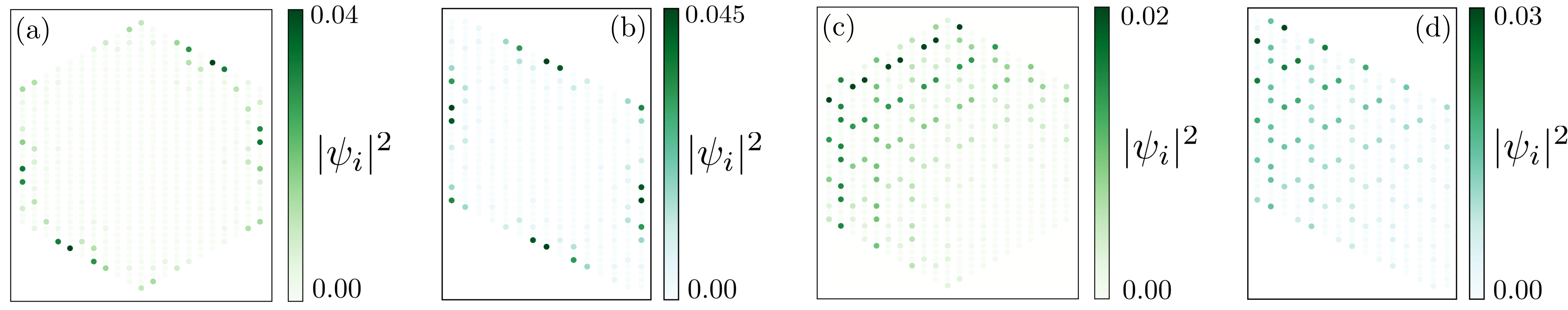}
\caption{Site-resolved normalized probability distributions ($|\psi_i|^2$) of the in-gap states apart from those contributing to MCMs in the intrinsic-SOTMI phase $J_0=0.7$ (topological regime) is shown in panels (a,c) and (b,d) for hexagonal and rhombus geometries respectively. Panels a,b represents $|\psi_i|^2$ of those in-gap states that are slightly below the states  contributing to MCMs. Panels c,d represents $|\psi_i|^2$ of those in-gap states that are in close proximity to the bulk states in the spectrum.  The lattice geometries in (a,c) and (b,d) consist of 222 sites (37 unit cells) and 150 sites (25 unit cells), respectively with type-1 termination. Other model parameters are chosen as $(J_1, \beta, S)=(1, \tfrac{1}{3}, \tfrac{3}{2})$.}\label{fig:Fig6}
\end{figure}
%-----------------------------------------------------------------------------------------

%\pagebreak
%%%%%%%%%%%%%%%%%%%%%%%%%%%%%%%%%%%%%%%%%%%%%%%%%%%%%%%%%%%%%%%%%%%%%%%%%%%%%
%\bibliographystyle{apsrev4-1}
%\bibliography{ref}
%%%%%%%%%%%%%%%%%%%%%%%%%%%%%%%%%%%%%%%%%%%%%%%%%%%%%%%%%%%%%%%%%%%%%%%%%%%%%

\end{onecolumngrid}

\end{document}